\documentclass[]{birkjour}
\usepackage[noadjust]{cite}
\usepackage{xcolor}
\RequirePackage[all]{xy}

\theoremstyle{definition}

\theoremstyle{remark}

\numberwithin{equation}{section}

\newcommand{\BibTeX}{B\kern-0.1emi\kern-0.017emb\kern-0.15em\TeX}
\newcommand{\XYpic}{$\mathrm{X\kern-0.3em\raisebox{-0.18em}{Y}}$-$\mathrm{pic}\,$}

\newcommand{\cl}{C \kern -0.1em \ell}  

\newcommand{\ed}{\end{document}}

\usepackage{amsfonts}
\usepackage{bbm} 
\usepackage{graphicx}

\begin{document}

%
%
%
%
%
%
%
%
%

\title[Dirac Quasinormal Modes in Noncommutative RN Black Holes]
 {Dirac Quasinormal Modes in Noncommutative Reissner–Nordström Black Holes}
\author[N. Herceg]{Nikola Herceg}
\address{%
Rudjer Bo\v{s}kovi\'c Institute, Bijeni\v cka  c.54, HR-10002 Zagreb, Croatia}
\email{Nikola.Herceg@irb.hr}
%
\author[N. Konjik]{Nikola Konjik}
\address{%
Faculty of Physics, University of Belgrade,  Studentski trg 12, 11000 Beograd, Serbia}
\email{konjik@ipb.ac.rs}

\author[A. Naveena Kumara]{A. Naveena Kumara}
\address{%
Rudjer Bo\v{s}kovi\'c Institute, Bijeni\v cka  c.54, HR-10002 Zagreb, Croatia}
\email{nathith@irb.hr}

\author[A. Samsarov]{Andjelo Samsarov}
\address{%
Rudjer Bo\v{s}kovi\'c Institute, Bijeni\v cka  c.54, HR-10002 Zagreb, Croatia}
\email{asamsarov@irb.hr}
\date{\today}
\dedicatory{Last Revised:\\ \today}
\begin{abstract}
Noncommutative (NC) geometry provides a novel approach to probe quantum gravity effects in black hole spacetimes. This work explores Dirac quasinormal modes (QNMs) of a deformed Reissner-Nordström black hole, where noncommutativity induces an effective metric with an additional ($ r-\varphi$) component. Employing a semiclassical model equivalent to a NC gauge theory, we investigate the dynamics of massless Dirac fields and calculate their QNM frequencies using the continued fraction method, enhanced by Gauss elimination to address the six-term recurrence relations. Our results demonstrate notable shifts in oscillation frequencies and damping rates relative to the commutative Reissner-Nordström case, exhibiting a distinctive Zeeman-like splitting in the QNM spectrum driven by the NC parameter.
\end{abstract}
\label{page:firstblob}
\maketitle
\tableofcontents

\section{Introduction}

Black holes, once confined to the realm of mathematical curiosity, have emerged as empirical laboratories for exploring gravity in its most extreme form. The detection of gravitational waves by LIGO, Virgo, and KAGRA has transformed these compact objects into astrophysical tools, opening a new observational frontier often referred to as gravitational-wave astronomy~\cite{Berti:2025hly}. Within this framework, the quasinormal modes (QNMs) of black holes—those damped oscillations that arise from perturbations—have become central observables. These modes, akin to the resonant frequencies of a bell, carry detailed information about the geometry of the black hole and the underlying gravitational theory. They offer a novel means to probe not only classical general relativity but also its possible extensions at the quantum level. Indeed, in the emerging field of gravitational spectroscopy, one of the key motivations is to investigate whether QNMs can resolve fine structures in the spacetime spectrum—structures that may reveal imprints of quantum gravity.

Among the diverse proposals for quantum gravity, noncommutative (NC) geometry provides a particularly elegant and mathematically rich framework. Inspired by developments in quantum field theory and string theory, NC geometry replaces the classical notion of spacetime with a NC algebra of functions, thereby introducing a minimal length scale~\cite{Seiberg:1999vs}. This “fuzziness” at the Planck scale leads to modifications in the local structure of spacetime that could, in principle, manifest in strong gravity regimes. In the semiclassical approach adopted here, one considers deformations of the spacetime metric induced by NC gauge field interactions while treating the gravitational background classically. A particularly tractable case is obtained when the NC deformation is encoded through a Drinfel'd twisting of the underlying symmetry algebra. Such deformations can be systematically implemented using the tools of twist-deformed differential geometry, leading to consistent field theories on NC curved backgrounds.


In this work, we explore the propagation of massless Dirac fields in a background of NC-deformed Reissner–Nordström (RN) black hole. 
The deformation leads to a twist-induced splitting of the fermionic QNM spectrum, reminiscent of the Zeeman effect in atomic physics. The analogy is striking and suggests that black holes, when viewed through the lens of NC geometry, may exhibit “quantum fingerprints” in their spectral response. Our approach draws further motivation from earlier studies of scalar and gravitational QNMs on NC black hole backgrounds, where similar splitting phenomena have been observed \cite{Herceg:2023pmc, Herceg:2023zlk, Herceg:2024vwc}.

The investigation is grounded in a semiclassical model where NC gauge and fermion fields interact with a classical RN geometry. By reducing the governing equations to a Schrödinger-like form and employing Leaver’s continued fraction method—extended here to handle six-term recurrence relations—we extract the QNM frequencies. The analysis reveals not only qualitative changes in the spectrum but also quantitative trends sensitive to the NC parameter, magnetic quantum number, and charge coupling. The results indicate that noncommutativity leaves an imprint on the fermionic QNM spectrum, offering a novel mechanism for probing Planck-scale effects in black hole physics.

\section{Dirac field in NC curved spacetime}

We begin by outlining the mathematical framework for describing the propagation of a Dirac field in a NC curved spacetime. For a detailed derivation, the reader is referred to Ref.~\cite{Herceg:2025zkk}. The NC deformation considered here is constructed via deformation quantization using the Drinfeld twist formalism. In particular, we employ the so-called \emph{angular twist} operator~\cite{Ciric:2017rnf, DimitrijevicCiric:2019hqq}:
\begin{eqnarray}  \label{AngTwist0Phi}
\mathcal{F} = e^{-\frac{i}{2}\theta ^{\alpha\beta}\partial_\alpha\otimes \partial_\beta} 
= e^{-\frac{ia}{2} (\partial_t\otimes\partial_\varphi - \partial_\varphi\otimes\partial_t)}, \nonumber
\end{eqnarray}
where $ \alpha,\beta \in \{t,r, \theta, \varphi\}$, and the only non-vanishing component of the deformation tensor is $\theta^{t\varphi} = -\theta^{\varphi t} = a$. The parameter $a$ characterizes the strength of noncommutativity and is typically associated with the Planck scale. As the twist is built from Killing vectors of the underlying spacetime, it qualifies as a \emph{Killing twist}.

It was shown in Ref.~\cite{DimitrijevicCiric:2022ohs} that, under this angular twist, the dynamics of twist-deformed $U(1)$ gauge theories can be recast into a commutative form, but in a modified spacetime geometry. Concretely, the equation of motion for
a charged NC scalar field with NC $U(1)$ gauge coupling propagating in a classical RN background,
\begin{equation*}
  {\rm d}s^2 = f\, {\rm d}t^2 - f^{-1} {\rm d}r^2   - r^2 {\rm d}\Omega^2_2 \quad , \text{with} \quad f(r) = 1 - \frac{2M}{r} + \frac{Q^2}{r^2},
\end{equation*}
is found to be equivalent to that of an ordinary (commutative) scalar field with the same charge $q$, but propagating in a modified RN geometry,
\begin{equation}\label{modi}
{\rm d}s^2 = f\, {\rm d}t^2 - f^{-1} {\rm d}r^2 - aqQ \sin^2 \theta\, {\rm d}r\, {\rm d}\varphi
  - r^2 {\rm d}\Omega^2_2.
\end{equation}
This modified metric differs from the classical RN solution by the presence of an off-diagonal $r$–$\varphi$ term, which arises solely due to spacetime noncommutativity and is non-zero only in the presence of a charged field.

Motivated by the duality arguments mentioned above, we consider the Dirac equation on the background defined by the NC-deformed RN metric given in Eq.~\eqref{modi}:
\begin{equation}
\big( i \gamma^a ( \nabla_a + A_a ) - m \big)\Psi = \big( i \gamma^a e_a^{~~\mu} ( \nabla_{\mu} + A_{\mu} ) - m \big)\Psi =0. \nonumber
\end{equation}
Here, $\gamma^a$ denote the flat spacetime Dirac gamma matrices, which satisfy the Clifford algebra $\{ \gamma_a, \gamma_b \} = 2 \eta_{ab}$, where $\eta_{ab} = \text{diag}(+1, -1, -1, -1)$. The curved-space Dirac operator $ \gamma^a \nabla_a$ is constructed using the tetrads $e^a_{~\mu}$, which relate the curved metric to the flat tangent space via the identity $g_{\mu \nu} = e^a_{~\mu} e^b_{~\nu} \eta_{ab}$. We adopt the following representation for the gamma matrices:
\begin{eqnarray}
    \label{gammarep}
  \gamma^0 &=& i  \tilde{\gamma}^0 =
 i \left( \begin{array}{cc}
  0  & I  \\
   I   & 0  \\ 
\end{array} \right),  \qquad 
 \gamma^1 =   i \tilde{\gamma}^3 =
i \left( \begin{array}{cc}
  0  & \sigma_3  \\
   -\sigma_3   & 0  \\  
 \end{array} \right), \nonumber \\
 \gamma^2 &=&   i \tilde{\gamma}^1 =
i\left( \begin{array}{cc}
  0  & \sigma_1  \\
   -\sigma_1   & 0  \\ 
\end{array} \right),  \quad 
 \gamma^3 =   i \tilde{\gamma}^2 =
i\left( \begin{array}{cc}
  0  & \sigma_2  \\
   -\sigma_2   & 0  \\ 
\end{array} \right),
\end{eqnarray}
where $\tilde{\gamma}^\mu$ are the gamma matrices in chiral basis, and $\sigma_i$ $(i = 1, 2, 3)$ denote the standard Pauli matrices~\cite{Dolan:2015eua, Richartz:2014jla}.

The spinor $\Psi$ is decomposed into two two-component spinors, $\Psi_1$ and $\Psi_2$, such that $\Psi = (\Psi_1, \Psi_2)^T$. The background gauge potential is taken to be $A_\mu = (A_t, \vec{A}) = \left(-\frac{qQ}{r}, \vec{0}\right)$. Substituting this into the Dirac equation yields a coupled system of equations:
\begin{equation}
\begin{pmatrix}
\mathcal D_+ & -m\,\mathbbm{1} \\
-m\,\mathbbm{1} & \mathcal D_-
\end{pmatrix}
\begin{pmatrix}\Psi_1 \\ \Psi_2\end{pmatrix}
=0\, , 
\end{equation}
where the differential operators $\mathcal D_\pm$ are given by
\begin{eqnarray}
\mathcal D_{\pm} &=& -\frac{1}{\sqrt f}\,\mathbbm{1}\,\partial_t
   \pm \sqrt f\,\sigma_3\,\partial_r
   \pm \frac{1}{2}\,\frac{Mr-Q^2}{r^3}\,\frac{1}{\sqrt f}\,\sigma_3
   \pm \frac{\sqrt f}{r}\,\sigma_3 
   \label{eq:D-line1} \nonumber\\
&& \pm \;\frac{1}{r}\,\sigma_1\,\partial_\theta
   \mp \frac{a q Q}{2r^2}\,\sqrt f\,\sigma_3\,\partial_\varphi
   \pm \frac{1}{r\sin\theta}\,\sigma_2\,\partial_\varphi
   \pm \frac{1}{2r}\,\cot\theta\,\sigma_1
   -\frac{i q Q}{r\sqrt f}\,\mathbbm{1}\,. \nonumber
   \label{eq:D-line2}
\end{eqnarray}
To achieve separation of variables, we adopt the ansatz
\begin{eqnarray}\label{anz1}
\Psi &=&
  e^{i(\nu \varphi - \omega t)} \left( \psi_1 (r, \theta),  \psi_2 (r, \theta ) \right)^T \\
  &=&
   e^{i(\nu \varphi - \omega t)}  \left( p R_2 S_1,\  - \frac{1}{r} R_1 S_2,\  \frac{1}{r} R_1 S_1,\  R_2 S_2 \right)^T,
\end{eqnarray}
where $R_{1,2}(r)$ and $S_{1,2}(\theta)$ are radial and angular functions, respectively, and $p \in \{-1, +1\}$ is a parameter used to classify solutions. This leads to a pair of independent angular equations:
\begin{equation}
\left[\,\partial_\theta+\frac12\cot\theta-\frac{\nu}{\sin\theta}\,\sigma_3\,\right]\,S(\theta)
=\Lambda\,S(\theta),
\qquad 
S(\theta)=\begin{pmatrix}S_1\\[2pt]S_2\end{pmatrix},\quad
\Lambda=\begin{pmatrix}0&\lambda_1\\[2pt]\lambda&0\end{pmatrix}, \nonumber
\end{equation}
along with two coupled radial equations:
\begin{equation}
\begin{pmatrix}
\mathcal{R}_+ & -(\lambda+p m r)\\[4pt]
-(p\lambda - m r) & r^2 \mathcal{R}_- + r\sqrt{f}
\end{pmatrix}
\begin{pmatrix}R_1\\[2pt]R_2\end{pmatrix}=0,
\end{equation}
where the differential operators $\mathcal{R}_\pm$ are defined as
\[
\mathcal{R}_\pm = \frac{i\omega}{\sqrt{f}}
                     \mp \sqrt{f}\,\partial_r
                     \mp \frac{1}{2}\frac{Mr-Q^{2}}{r^{3}}\frac{1}{\sqrt{f}}
                     \pm i\nu\frac{a q Q}{2 r^{2}}\sqrt{f}
                     -\frac{i q Q}{r\sqrt{f}}.
\]

Upon decoupling, the system reduces to two second-order radial equations of the form:
\begin{equation}
\mathcal{L}_{r1,r2} \,R_{1,2}(r) = (\lambda + p m r)\,(p\lambda - m r)\,R_{1,2}(r), 
\end{equation}
with the second-order differential operator
\begin{equation}
\mathcal{L}_{r1,r2} = -r^2 f(r)\,\partial_r^2 + A_{1,2}(r) \,\partial_r + B_{1,2}(r),
\end{equation}
corresponding to $R_1 \equiv R_{s=-1/2}$ and $R_2 \equiv R_{s=+1/2}$, where $s$ denotes the chirality. (The explicit forms of $A_{1,2}(r)$ and $B_{1,2}(r)$ can be found in Ref.~\cite{Herceg:2025zkk}.) Separating the Dirac spinor into chiral parts imposes the relation $\lambda^2 = (j-s)(j+s+1)$ \cite{Dolan:2015eua}.
By introducing the transformation $R_s = r^{1/2} \Delta^{-1} \xi_s$, where $\Delta = r^2 f(r)$, the radial equation simplifies to the form~\footnote{We adopt $p = -1$, which is the only physically consistent choice for massive spinor fields; it also holds in the massless limit. For explicit expressions of $B(r)$ and $C(r)$, see Ref.~\cite{Herceg:2025zkk}.}
\[
\left[\,
\Delta\,\partial_r^2
\;+\;B(r)\,\partial _r
\;+\;C(r)
\right]\,\xi_s(r) = 0.
\]

In this work, we restrict ourselves to the massless case $m = 0$. Under this assumption, the radial equation can be further cast into a Schrödinger-like form:
\begin{equation}
\frac{{\rm d}^2 \chi}{{\rm d} y^2} + V\,\chi = 0,
\end{equation}
where the tortoise coordinate $y$ is defined via
\[
\frac{{\rm d}y}{{\rm d}r} = \frac{1}{f(r)\left(1 + i a \nu \frac{qQ}{r}\right)},
\]
and the field redefinition $\chi_s(r) = \Delta^{s/2} r\, \xi_s(r)$ has been applied.
The resulting effective potential $V$ takes the form
\begin{eqnarray} \label{veffektive}
V &=& \frac{\Delta}{r^4} \Bigg[ \frac{2Q^2}{r^2} - \frac{2M}{r} - j(j+1) + s^2 
    + \frac{\left( \omega r^2 - qQr - i s(r - M) \right)^2}{\Delta} \nonumber \\
    &+& 4i s \omega r - 2i s qQ
    + \frac{i a \nu qQ \Delta}{r^3} + i s a \nu qQ \frac{r - M}{r^2} \nonumber \\
    &-& \frac{i a \nu qQ}{r^3} \left( s r^2 + (1-s) M r - Q^2 \right)
    + 2i a \nu \frac{qQ}{r} \left( \frac{2Q^2}{r^2} - \frac{2M}{r} - j(j+1) + s^2 \right) \nonumber \\
    &+& 2i a \nu \frac{qQ}{r} \frac{\left( \omega r^2 - qQr - i s(r - M) \right)^2}{\Delta}
    - 8s a \nu \omega qQ + 4s a \nu \frac{q^2 Q^2}{r} \Bigg]. \nonumber
\end{eqnarray}
This expression is valid up to linear order in the NC deformation parameter $a$, and clearly illustrates how noncommutativity modifies both the structure and complexity of the effective potential.

\section{NC fermion quasinormal modes}

In this section, we compute the QNM spectrum of the massless Dirac field in the NC background using Leaver’s continued fraction method. As a first step, we analyze the behavior of the radial equation near the black hole horizon and at spatial infinity. The standard boundary conditions for QNMs apply: the solution must be purely ingoing at the event horizon and purely outgoing at infinity. Accordingly, the asymptotic form of the solutions is given by:
\begin{gather}    
  \xi_s (r) \rightarrow 
\begin{cases}         
  Z_{\text{out}}\, e^{i \omega y}\, y^{-1 - i qQ - 2s - a\nu qQ \omega}, & \text{as } r \rightarrow \infty , \\[10pt]
  Z_{\text{in}}\, {(r - r_+)}^{-s/2} \, e^{-i \left( \omega  - \frac{qQ}{r_+} - i s \frac{r_+ - r_-}{2r_+^2} \right)\left( 1 + i a \nu  \frac{qQ}{r_+} \right)y}, & \text{as } r \rightarrow r_+ ,
\end{cases} \nonumber
\end{gather}
where $Z_{\text{out}}$ and $Z_{\text{in}}$ denote the amplitudes of the outgoing and ingoing modes, respectively.

The radial differential equation possesses an irregular singularity at $r = \infty$, and regular singularities at $r = 0$, $r = r_-$, and $r = r_+$. Following Leaver’s approach~\cite{Leaver:1990zz}, we adopt the following ansatz for the solution:
\begin{eqnarray}
 \xi_s (r) = e^{i \omega r}  (r - r_-)^{\epsilon} \sum_{n=0}^{\infty} a_n \left( \frac{r - r_+}{r - r_-} \right)^{n + \delta}.
\end{eqnarray}
Here, the parameters $\epsilon$ and $\delta$ are determined by matching the behavior of the series with the asymptotic solutions, and they remain unaffected by the NC deformation.

Substituting the above series into the radial equation yields a six-term recurrence relation for the expansion coefficients $a_n$:
\begin{eqnarray}  \label{6contfr}
A_n a_{n+1} + B_n a_n + C_n a_{n-1} + D_n a_{n-2} + E_n a_{n-3} + F_n a_{n-4} &=& 0, \quad  n \geq 4, \nonumber \\
A_3 a_{4} + B_3 a_3 + C_3 a_{2} + D_3 a_{1} + E_3 a_{0} &=& 0, \quad n = 3, \nonumber \\
A_2 a_{3} + B_2 a_2 + C_2 a_{1} + D_2 a_{0} &=& 0, \quad n = 2, \nonumber \\
A_1 a_{2} + B_1 a_1 + C_1 a_{0} &=& 0, \quad n = 1, \nonumber \\
A_0 a_{1} + B_0 a_0 &=& 0, \quad n = 0. \nonumber
\end{eqnarray}
This extended recurrence relation arises due to the presence of the NC correction terms. In contrast to the commutative case, where a three-term recurrence relation is obtained, the NC setup introduces additional complexity that is best addressed numerically.

The explicit expressions for the recurrence coefficients $A_n, B_n, C_n, D_n, E_n$, and $F_n$ in the six-term relation are given below:
\begin{eqnarray}  \label{contfr1}
  A_n   &=&   r_+^3 \alpha_{n},  \nonumber \\
  B_n  &=&  r_+^3 \beta_n - 3 r_+^2 r_- \alpha_{n-1}   -ia\nu qQr_+\Big[\frac{r_+-r_-}{2}+(n-s)(r_+-r_-) \nonumber\\
  && -ir_+(\omega r_+-qQ)+(r_+-r_-)\frac{s}{2}\Big], \nonumber \\
  C_n  &=&  r_+^3 \gamma_n  + 3r_+ r_-^2 \alpha_{n-2}  -3r_+^2 r_- \beta_{n-1}+a\nu qQ\omega r_+(r_+-r_-)^3 \nonumber\\
  &&+ia\nu qQ\Big[ (r_+-r_-)\big[(r_+-r_-)^2+\frac{r_-}{2}(r_+-r_-)\big]\nonumber\\
 &&-(r_+-r_-)^2[-1-2s-iqQ+i\omega (r_++r_-)] r_+ \nonumber \\
 &&+(r_+-r_-)(2r_++r_-)\big[(n-1-s)(r_+-r_-)-ir_+(\omega r_+-qQ)\big] \nonumber \\
 &&+sr_+(r_+-r_-)(r_++2r_-)-s(r_+-r_-)(2r_++r_-)\frac{r_++r_-}{2} \Big],\nonumber \\ 
D_n  &= &- r_-^3 \alpha_{n-3}  + 3r_+ r_-^2 \beta_{n-2} -3 r_+^2 r_- \gamma_{n-1}+ia\nu qQ\Big[ \frac{1}{2}r_+(r_+-r_-)^2\nonumber\\
&&-(r_+-r_-)(2r_++r_-)\big((n-2-s)(r_+-r_-)-ir_+(\omega r_+-qQ)\big)\nonumber\\
&&+(r_+-r_-)^2(r_++r_-)\big(-1-2s-iqQ+i\omega (r_++r_-)\big)\nonumber\\
&&-(r_+-r_-)^3+\frac{1}{2}s(r_+-r_-)(r_++2r_-)(r_++r_-)\nonumber \\
&&-sr_-(r_+-r_-)(2r_++r_-)\Big]-a\nu qQr_-\omega (r_+-r_-)^3 ,  \nonumber\\
  E_n  &=&   3r_+ r_-^2 \gamma_{n-2} - r_-^3 \beta_{n-3} +ia\nu qQ \Big[ -(r_+-r_-)^2\frac{r_-}{2}\nonumber\\
&&-(r_+-r_-)^2\big(-1-2s-iqQ+i\omega (r_++r_-)\big)r_-\nonumber\\
&&+(r_+-r_-)r_-\big((n-3-s)(r_+-r_-)-ir_+(\omega r_+-qQ)\big)\nonumber\\
&&-sr_-(r_+-r_-)(2r_++r_-)+\frac{1}{2}s(r_+-r_-)(r_++2r_-)(r_++r_-)\Big],
 \nonumber \\
  F_n &=& -r_-^3 \gamma_{n-3}.\nonumber
\end{eqnarray} 
Here, the functions $\alpha_n$, $\beta_n$, and $\gamma_n$ encode the commutative part of the recurrence relation and are given by~\cite{Richartz:2014jla}:
\begin{eqnarray}
  \alpha_n  &=&  -(n+1) \Big( r_-(n-s+1)+r_+(-n+s-1-2iqQ+2ir_+\omega) \Big), \label{contfrsimple} \nonumber \\
    \beta_n  &=&   -r_+\Big(\lambda_s+2n^2-4ir_+\omega(2n+1+3iqQ)+6inqQ+2n-4(qQ)^2 \nonumber\\
    &&+3iqQ-8r_+^2\omega^2+s+1 \Big) +r_- \big(\lambda_s+2n(n+1+iqQ)+iqQ+s+1 \big)\nonumber\\
  &&-2i(2n+1)r_+r_-\omega, \nonumber \\
  \gamma_n &=& -\Big(n+2i\big(qQ-\omega(r_++r_-)\big)\Big)\Big(n(r_--r_+)+ir_+(-2qQ+2r_+\omega+is)+r_-s\Big). \nonumber
\end{eqnarray}

Using Gaussian elimination, the six-term recurrence relation can be reduced to a three-term one of the form~\cite{DimitrijevicCiric:2019hqq}:
\begin{align}
\tilde{\alpha}_0 a_{1} +\tilde{\beta}_0 a_{0} &= 0, \nonumber\\
\tilde{\alpha}_n a_{n+1} +\tilde{\beta}_n a_{n}+ \tilde{\gamma}_n a_{n-1} &=0, \qquad \text{for } n \geq 1. \nonumber
\end{align}
Due to the recursive nature of the elimination process, closed-form expressions for the new coefficients $\tilde{\alpha}_n$, $\tilde{\beta}_n$, and $\tilde{\gamma}_n$ are tedious. To speed up the convergence, they were computed numerically at each step.

\begin{figure}[t]
\centering
\includegraphics[width=0.45\textwidth]{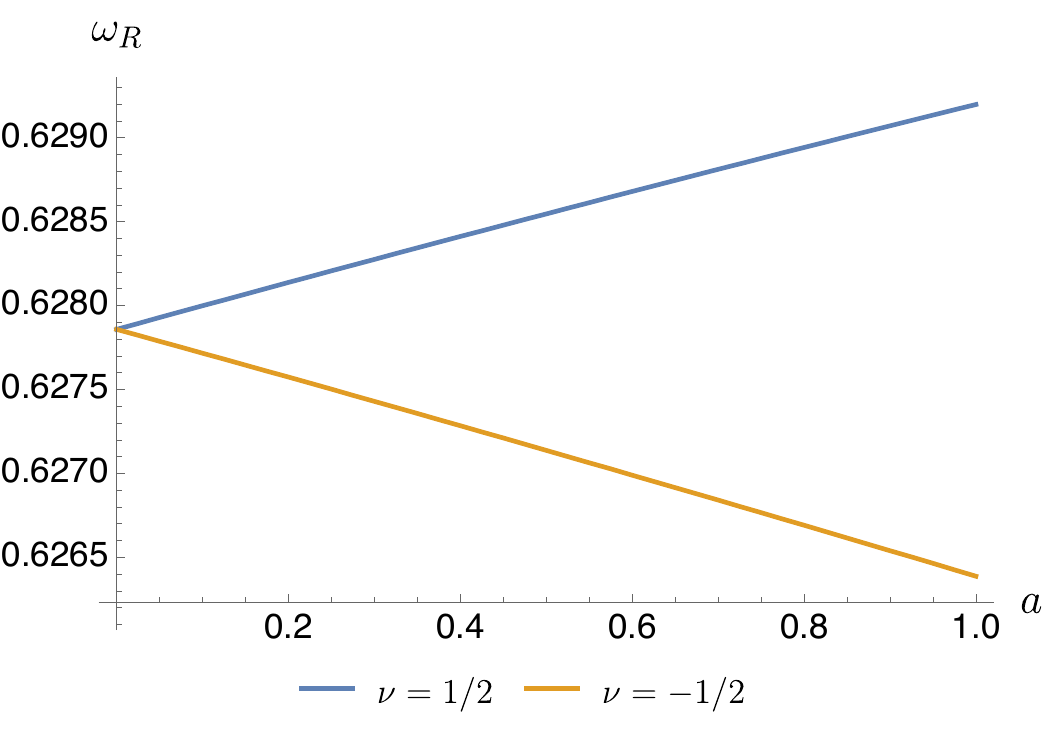}
\qquad
\includegraphics[width=0.45\textwidth]{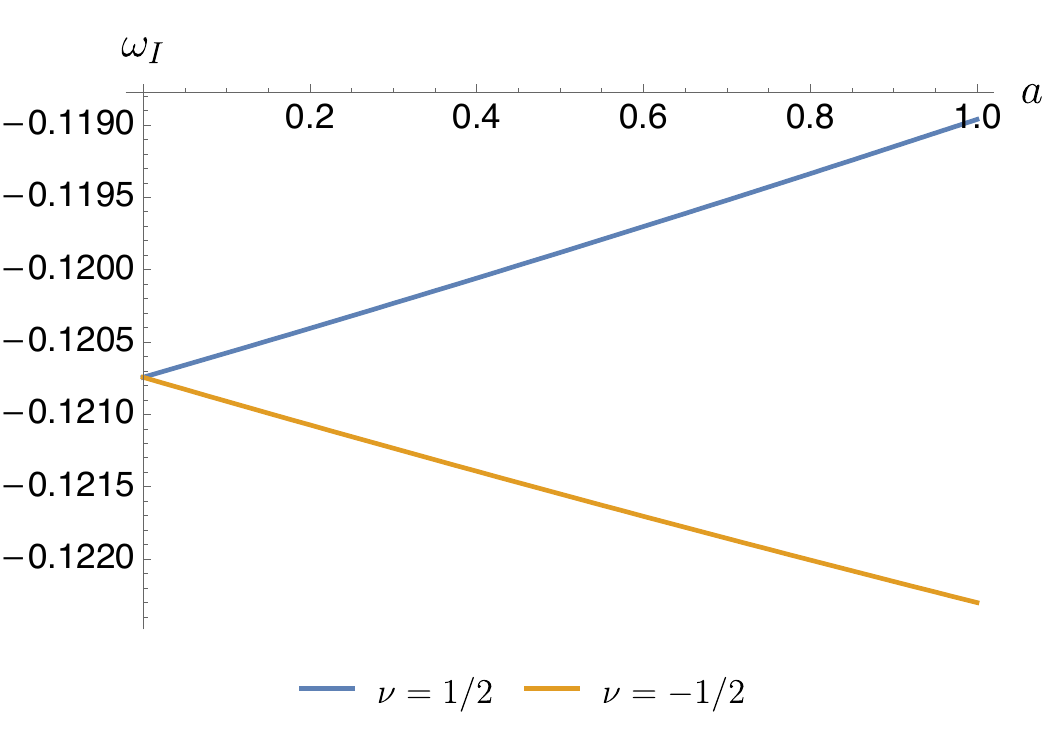}
\qquad
\includegraphics[width=0.45\textwidth]{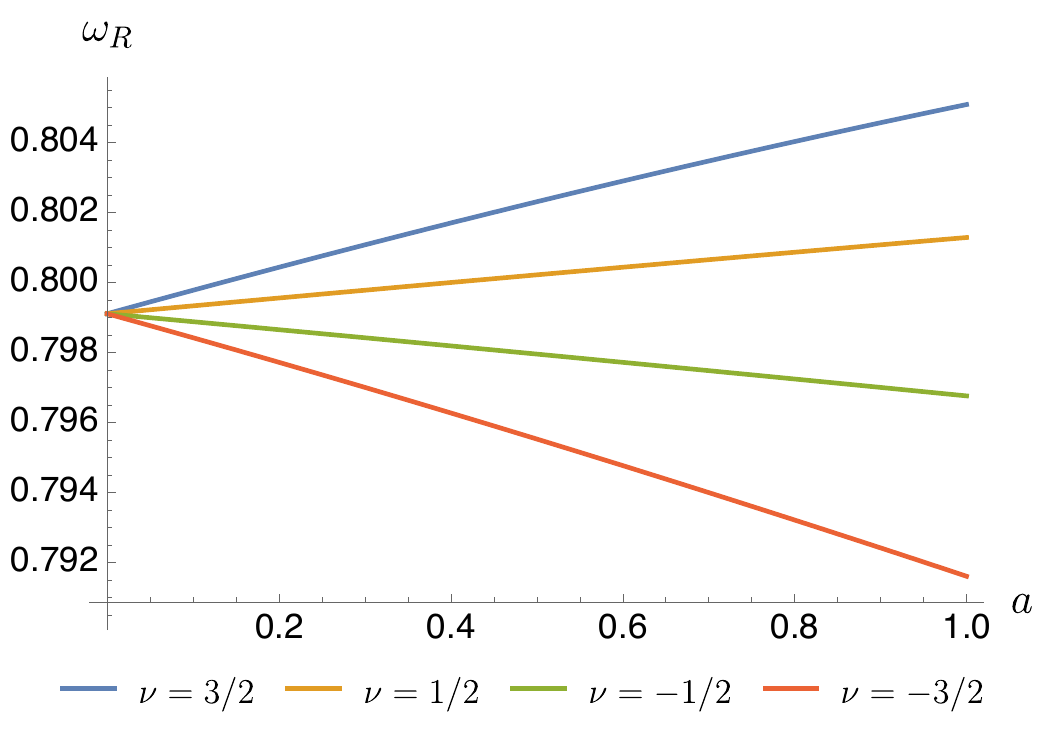}
\qquad
\includegraphics[width=0.45\textwidth]{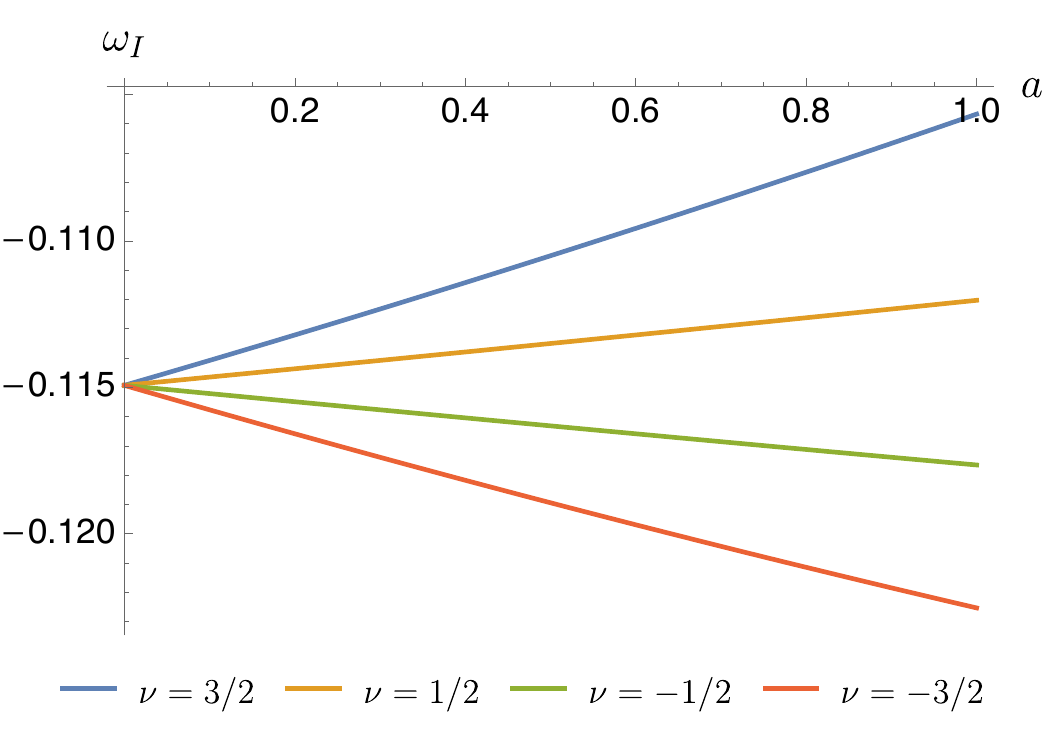}
\caption{\scriptsize QNM splitting for different values of the magnetic quantum number $\nu$. Top panel: $j = 1/2$, $s = 1/2$; Bottom panel: $j = 3/2$, $s = 1/2$.}\label{fig1}
\end{figure}

\begin{figure}[t]
\centering
\includegraphics[width=0.45\textwidth]{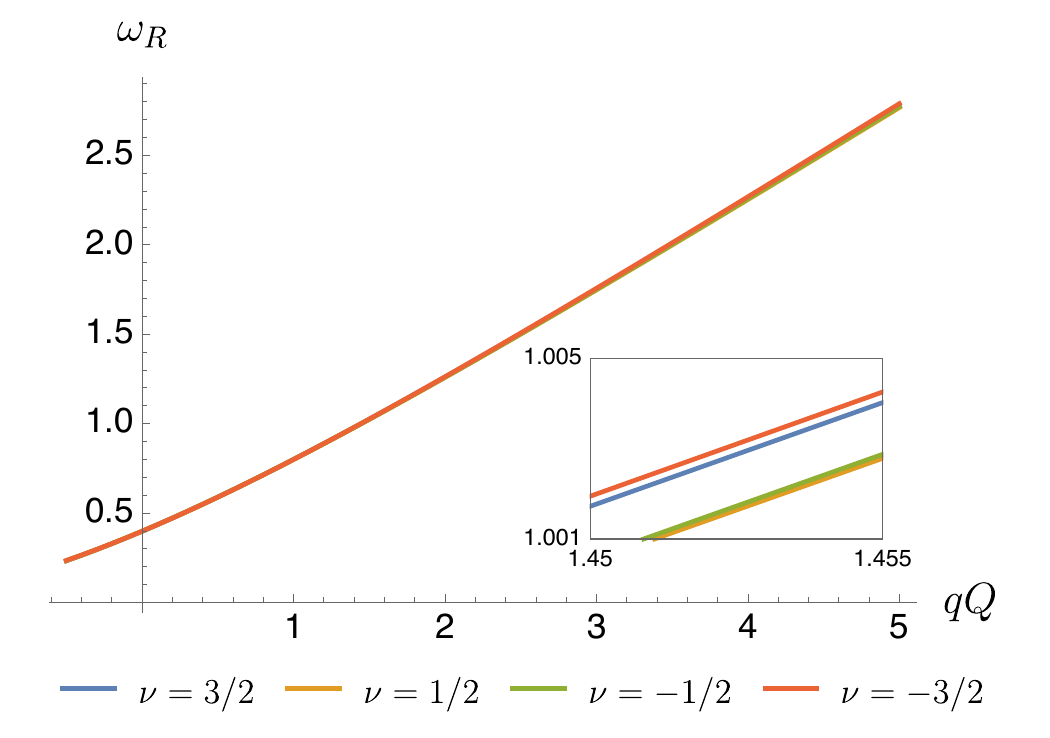}
\qquad
\includegraphics[width=0.45\textwidth]{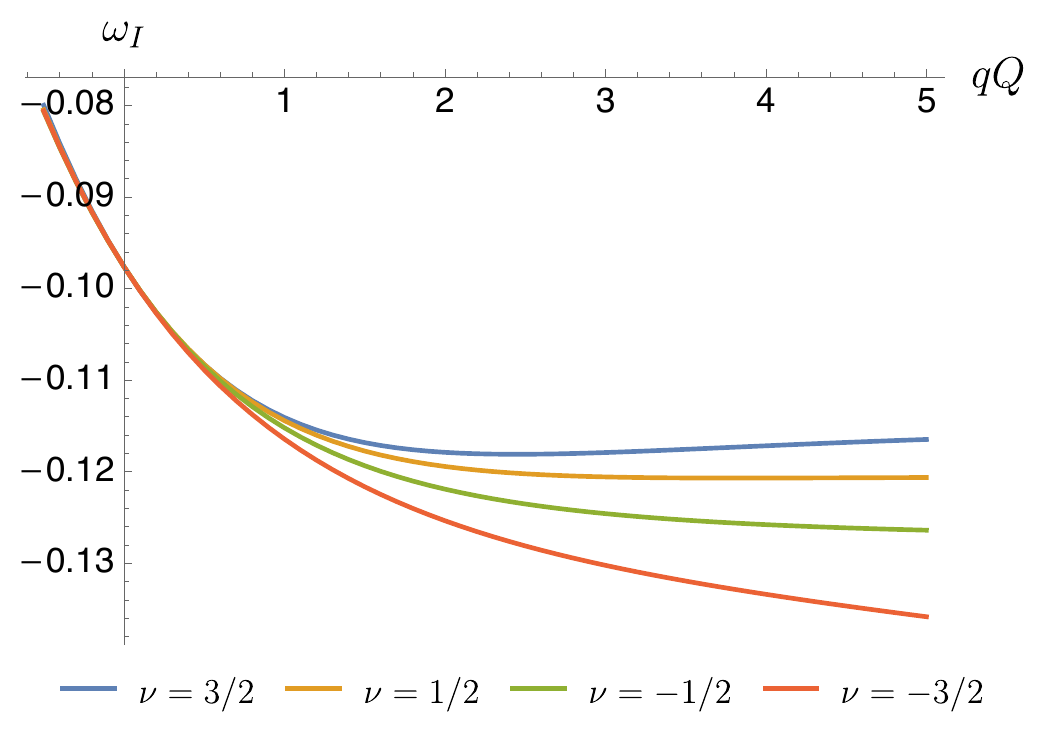}
\qquad
\includegraphics[width=0.45\textwidth]{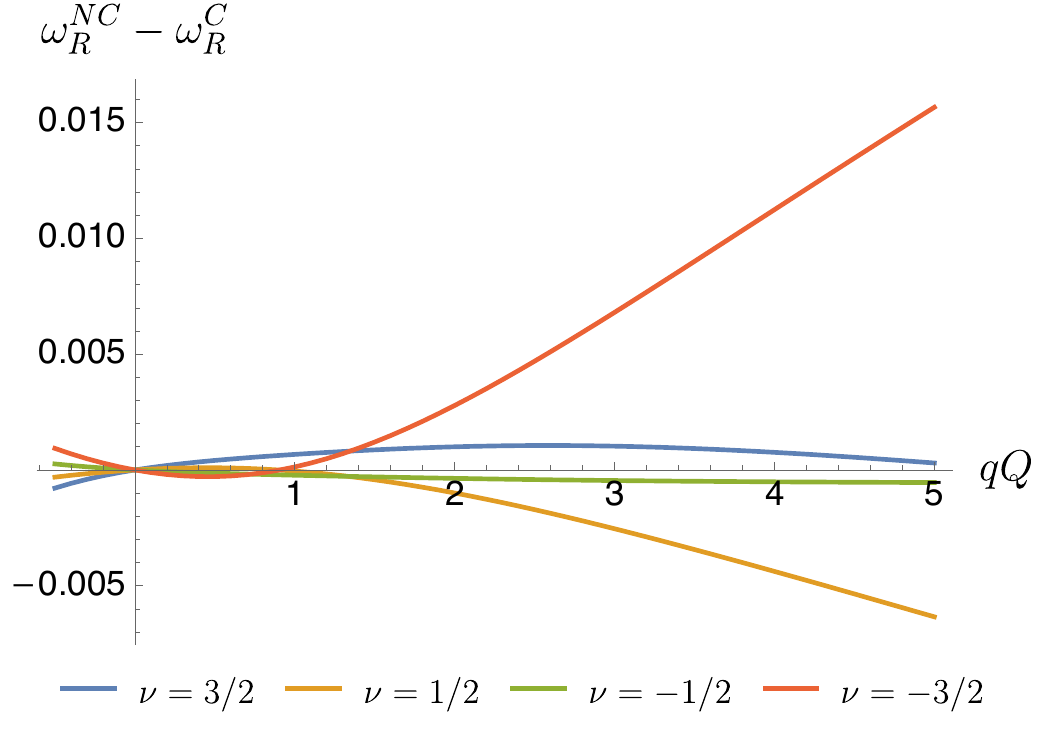}
\qquad
\includegraphics[width=0.45\textwidth]{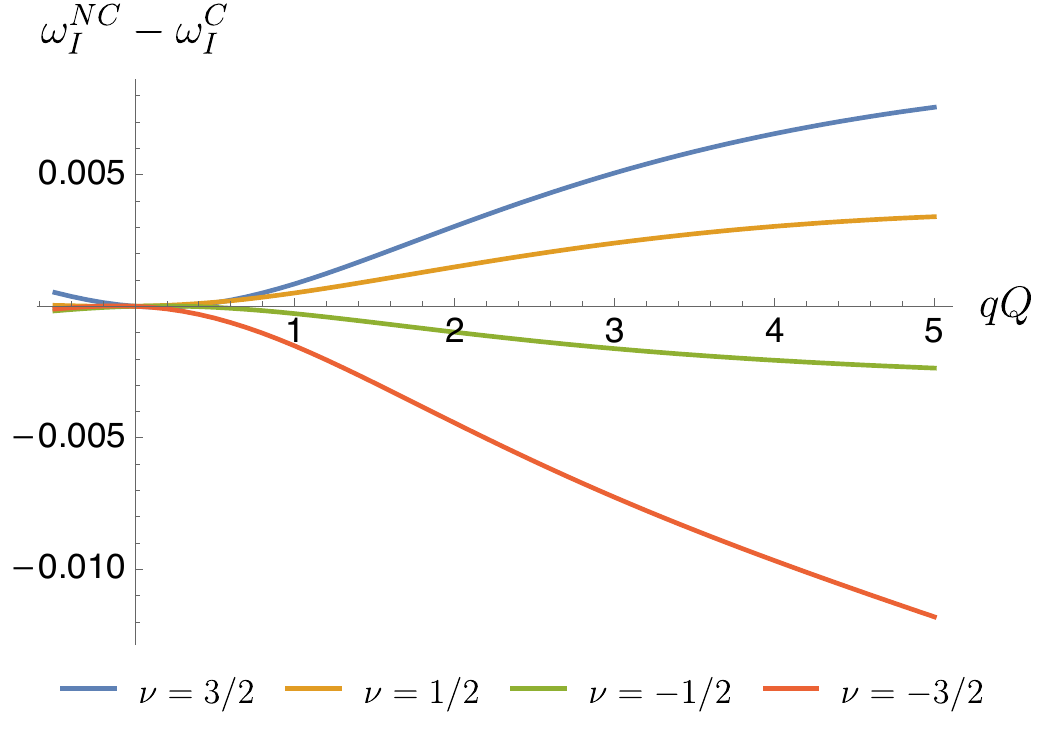}
\caption{\scriptsize Dependence of fermionic QNMs on $qQ$ for $j = 3/2$, $s = 1/2$.} \label{fig2}
\end{figure}

\begin{figure}[t]
\centering
\includegraphics[width=0.45\textwidth]{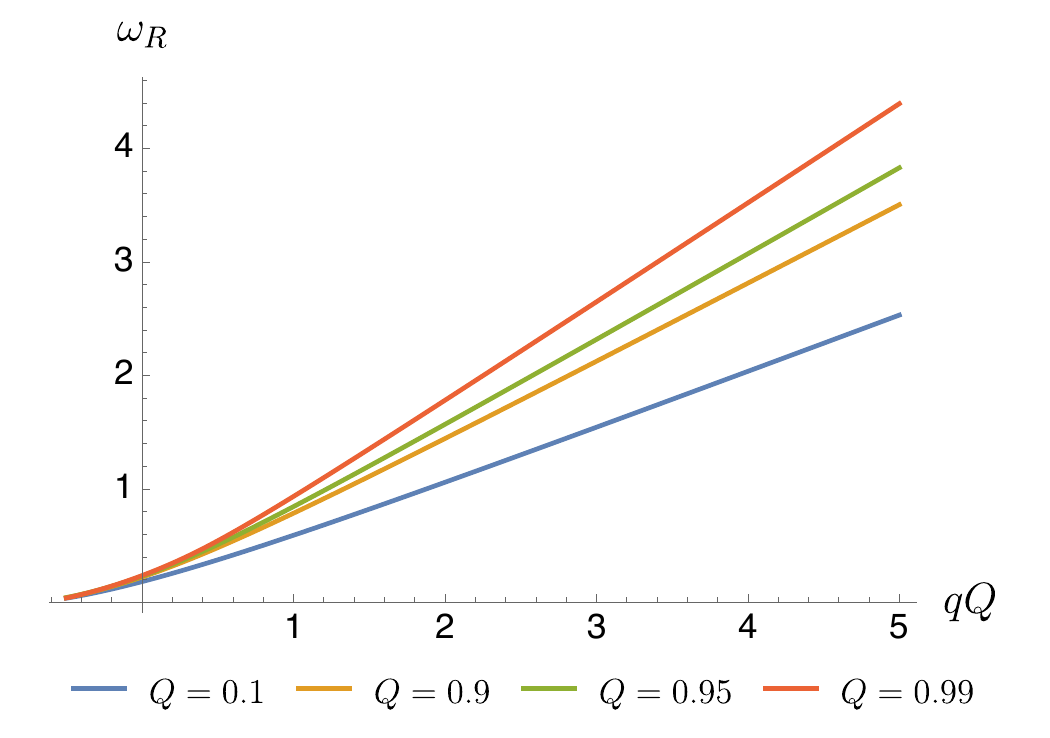}
\qquad
\includegraphics[width=0.45\textwidth]{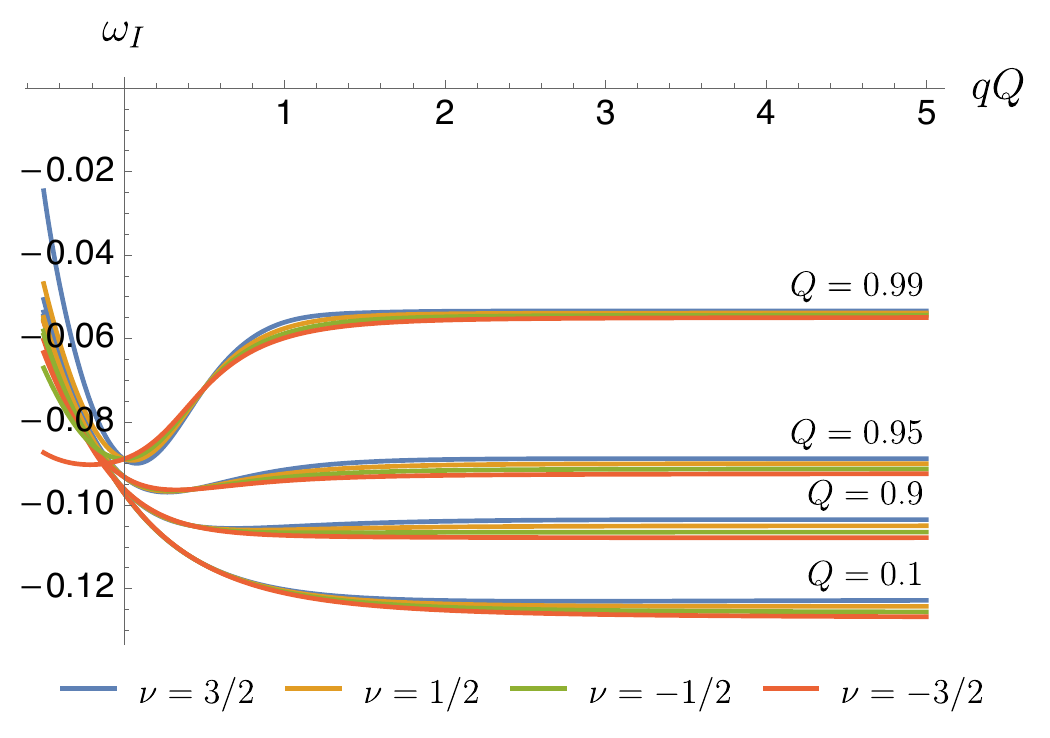}
\caption{\tiny Dependence of fermionic QNMs ($j = 3/2$, $s = 1/2$) on $qQ$.} \label{fig3}
\end{figure}

\begin{figure}[t]
\centering
\includegraphics[width=0.45\textwidth]{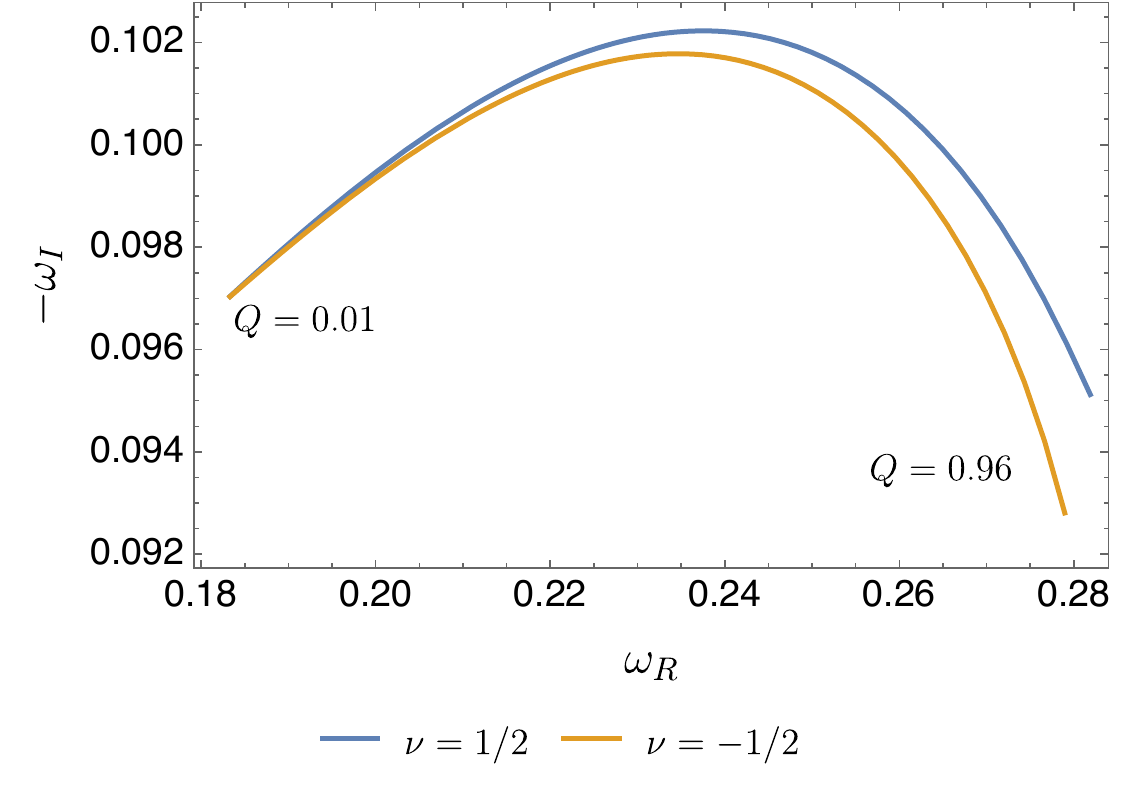}
\qquad
\includegraphics[width=0.45\textwidth]{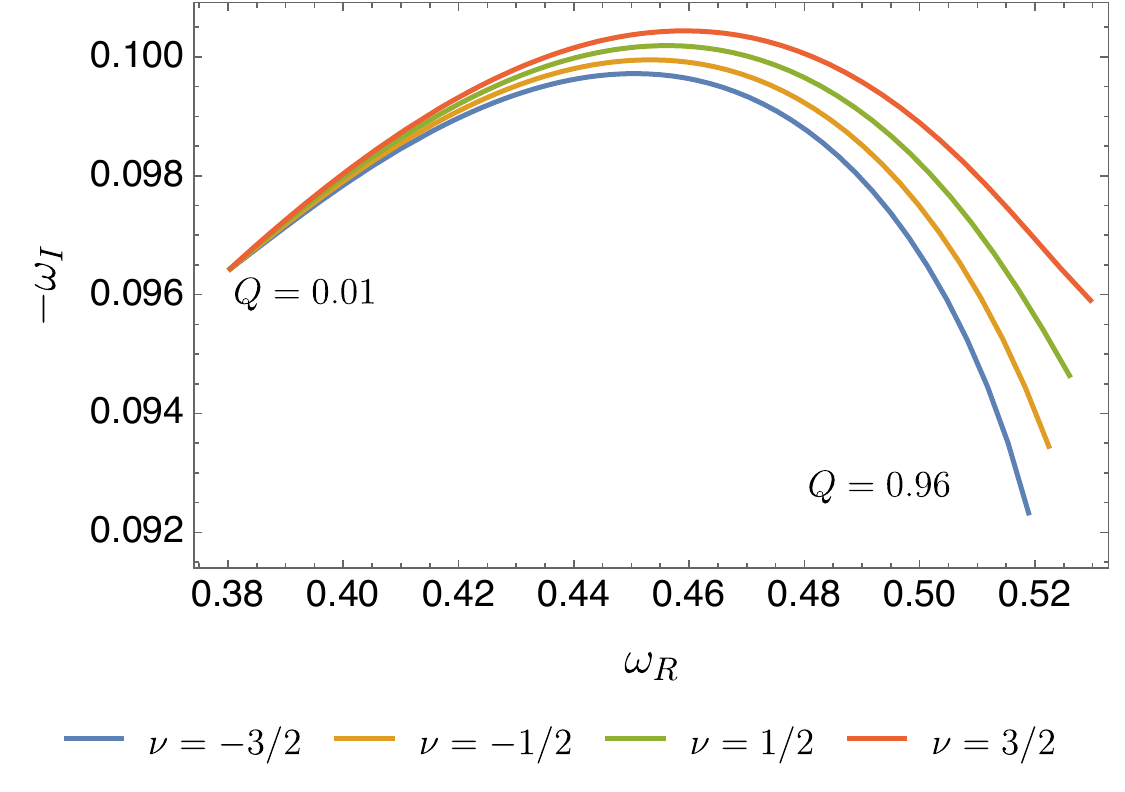}
\caption{\tiny $\omega_R$–$\omega_I$ portrait of QNMs.} \label{fig5}
\end{figure}

To ensure convergence of the series $\sum_n a_n$, these coefficients must satisfy the following continued fraction condition:
\[
\tilde{\beta}_0 = \frac{\tilde{\alpha}_0 \tilde{\gamma}_{1}}{\tilde{\beta}_{1}-\frac{\tilde{\alpha}_{1}\tilde{\gamma}_{2}}{\tilde{\beta}_{2} - \cdots}} = \frac{\tilde{\alpha}_0\tilde{\gamma}_{1}}{\tilde{\beta}_{1}-}\frac{\tilde{\alpha}_{1}\tilde{\gamma}_{2}}{\tilde{\beta}_{2} -}\frac{\tilde{\alpha}_{2}\tilde{\gamma}_{3}}{\tilde{\beta}_{3} -}\cdots 
\]
The QNM frequencies are then obtained by numerically locating the roots of this continued fraction, typically truncated after a finite number of terms (e.g., $N \sim 200$). To enhance numerical stability and improve precision, we employ Nollert’s method~\cite{Nollert:1993zz} to estimate the tail contributions of the truncated series. The results of this analysis are visualized in Figures~\ref{fig1}–\ref{fig5}.

Figure~\ref{fig1} illustrates the dependence of the QNM frequency on the NC parameter $a$. A clear linear deviation from the commutative case is observed, indicating a Zeeman-like splitting. This phenomenon mirrors what has been reported in scalar and spin-2 field studies~\cite{DimitrijevicCiric:2019hqq, Herceg:2023pmc, Herceg:2023zlk, Herceg:2024vwc}. It is worth noting that although the NC parameter $a$ is expected to be of the order of the Planck length—significantly smaller than unity in units where the black hole mass is normalized to $M = 1$—the values chosen here serve illustrative purposes only.

In Figure~\ref{fig2}, we explore the QNM spectrum as a function of the charge coupling $qQ$, keeping the black hole charge $Q$ fixed and considering $j = 3/2$. The plots reveal that NC effects significantly influence the imaginary part of the frequency $\omega_I$ (damping rate), especially for large values of $qQ$, while the real part $\omega_R$ remains comparatively less affected. Furthermore, the frequency differences between modes with $\nu = \pm 1/2, \pm 3/2$ highlight the twist-induced breaking of azimuthal symmetry—a manifestation of the preferred $\varphi$-direction introduced by noncommutativity. Although the modification to the metric is not due to physical rotation, the observed splitting in $\omega_I$ draws analogies with rotational effects~\cite{Leaver:1985ax, Detweiler:1980gk}.

Figure~\ref{fig3} presents the QNM spectrum as a function of $qQ$ for different values of $Q$. Notably, we observe a phase-transition-like feature in $\omega_I$ around $Q/M \sim 0.9$, consistent with the commutative case~\cite{Richartz:2014jla}. This characteristic behavior appears largely unaffected by the NC deformation.

Finally, Figure~\ref{fig5} displays the QNM trajectories in the $\omega_R$–$\omega_I$ plane, parametrized by the charge ratio $Q/M$. These plots reveal that the impact of noncommutativity becomes more pronounced as $Q/M$ increases. However, it is important to note that the current QNM computation technique breaks down in the extremal limit ($Q/M \rightarrow 1$), necessitating a separate analytical or numerical treatment in that regime.

\section{Concluding Remarks}

In this work, we investigated the propagation of a Dirac field in a NC-deformed RN black hole background and computed the corresponding QNM frequencies using Leaver’s continued fraction method. Our analysis was perturbative in the NC parameter $a$ and is valid to linear order in $a$.

The results highlight that NC geometry introduces distinct signatures in the QNM spectrum, most notably a Zeeman-like splitting of frequencies. Such effects suggest that QNMs can serve as potential probes of spacetime noncommutativity, enriching the framework of gravitational spectroscopy. 

Our study focused on massless fermions and non-extremal black holes. Future investigations should consider extensions to massive Dirac fields and extremal black hole limits, which require separate analytical treatment due to the breakdown of the current method in those regimes.


\subsection*{Acknowledgment}
This  research was supported by the Croatian Science Foundation Project No. IP-2020-02-9614 {\it{Search for Quantum spacetime in Black Hole QNM spectrum and Gamma Ray Bursts. The work of N.K.  is supported by Project 451-03-136/2025-03/200162 of the Serbian Ministry of Science, Technological Development and Innovation. }}

\bibliographystyle{spmpsci}
\bibliography{myBibLib} 

\end{document}